\documentclass[11pt]{article}

\usepackage{amsmath}
\usepackage{amssymb}
\usepackage{latexsym}
\usepackage{color}
\usepackage{dsfont}
\usepackage{graphicx}
\usepackage{appendix}

\def\be{\begin{eqnarray}}
\def\ee{\end{eqnarray}}

\def\b*{\begin{eqnarray*}}
\def\e*{\end{eqnarray*}}

\newtheorem{Theorem}{Theorem}[part]

\newtheorem{Proposition}{Proposition}[part]

\newtheorem{Assumption}{Assumption}[part]

\newtheorem{Remark}{Remark}[part]

\makeatletter \@addtoreset{equation}{section}

\@addtoreset{Definition}{section}

\@addtoreset{Theorem}{section}

\@addtoreset{Proposition}{section}

\@addtoreset{Property}{section}

\@addtoreset{Assumption}{section}

\@addtoreset{Corollary}{section}

\@addtoreset{Lemma}{section}

\@addtoreset{Remark}{section}

\@addtoreset{Example}{section}

\def \E{\mathbb{E}}
\def \F{\mathbb{F}}

\def \P{\mathbb{P}}

\def \R{\mathbb{R}}

\def\Ac{{\cal A}}

\def\Fc{{\cal F}}

\def\={\;=\;}
\def\.{\;.}

\def\eps{\varepsilon}

\def\reff#1{{\rm(\ref{#1})}}

\def\1{{\bf 1}}
\def \ep{\hbox{ }\hfill{ ${\cal t}$~\hspace{-5.1mm}~${\cal u}$   } }

\def \proof{{\noindent \bf Proof. }}
\def \ep{\hbox{ }\hfill$\Box$}

 \def\normeL2#1{\left\|{#1}\right\|_{L^2}}
 \title{Large liquidity expansion of super-hedging costs}

 \author{Dylan {\sc Possamai} \footnote{CMAP, Ecole Polytechnique Paris, dylan.possamai@polytechnique.edu.}
 \and H. Mete {\sc Soner}\footnote{ETH (Swiss Federal Institute of Technology),
Zurich, hmsoner@ethz.ch. Research partly supported by the
European Research Council under the grant 228053-FiRM.}
      \and Nizar {\sc Touzi}\footnote{CMAP, Ecole Polytechnique Paris, nizar.touzi@polytechnique.edu.
      Research supported by the Chair {\it Financial Risks} of the {\it Risk Foundation} sponsored by Soci\'et\'e
             G\'en\'erale, the Chair {\it Derivatives of the Future} sponsored by the {F\'ed\'eration Bancaire Fran\c{c}aise}, and
             the Chair {\it Finance and Sustainable Development} sponsored by EDF and Calyon. }}
             
 \date{\today}

\begin{document}

 \maketitle

 \begin{abstract}
We consider a financial market with liquidity cost as in \c{C}etin,
Jarrow and Protter
\cite{cjp} where the supply function $\mathbf{S}^\eps(s,\nu)$
depends on a parameter $\eps\ge 0$ with $\mathbf{S}^0(s,\nu)=s$
corresponding to the perfect liquid situation. Using the PDE
characterization of \c{C}etin, Soner and Touzi
\cite{cst} of the super-hedging cost
of an option written on such a stock, we provide a Taylor expansion
of the super-hedging cost in powers of $\eps$.  In particular,
we explicitly compute the first term in the expansion for a
European Call option and give bounds for the order of the expansion for a European Digital Option.

\vspace{10mm}


\vspace{5mm}

\end{abstract}
\newpage

\section{Introduction}

The classical option pricing equation of Black \& Scholes is derived
under several simplifying assumptions.   The ``infinite''
liquidity of the underlying stock process is one of them.
In an attempt to understand the impact of liquidity, 
\c{C}etin, Jarrow, Protter  and collaborators \cite{cjp,cjpw,cr}
postulated the existence of a supply curve $\mathbf{S}(t,s,\nu)$
which is the price of a share of the stock when one wants to buy $\nu$
shares at time $t$.  In the Black \& Scholes setting, this price function is
taken to be independent of $\nu$ corresponding to infinite
amount of supply, hence infinite liquidity.  In a recent paper,
\c{C}etin, Soner and Touzi  \cite{cst} used this model and studied
the liquidity premium in the price of an option written on such
a stock with less than infinite liquidity. 
They characterized the option price by a nonlinear
Black \& Scholes equation, given in \reff{PDE} below.
In this pricing equation the
liquidity manifests itself
by means of a {\em liquidity function} $\ell$,
which is given by
\begin{equation*}
 \ell(t,s)
 :=
 \left[ 4 \ \frac{\partial\mathbf{S}}{\partial\nu}(t,s,0)\right]^{-1},\ \
 (t,s)\in[0,T]\times\R_+.
\end{equation*}
 The liquidity function $\ell$ measures the level of liquidity of the market.
 Namely, the larger $\ell$ is, the more liquid the market is.

The main result of \cite{cst} is the characterization
of the liquidity premium as the unique viscosity solution
of a nonlinear Black-Scholes equation \reff{PDE}, which is
very similar to the one derived by Barles and Soner
\cite{barles-soner}. This nonlinear equation
can only be solved numerically as no explicit solutions
are available. Motivated by this fact, in this paper
we obtain rigorous asymptotic expansions for the
liquidity premium. For vanilla options with sufficiently regular payoff, this expansion
can be calculated explicitly giving further insight into the
liquidity effects.

\noindent As stated the chief objective of this paper is to analyze the
large liquidity effect. Thus, we assume that the supply function
 depends on a  small parameter $\epsilon$
\begin{equation*}
\mathbf{S}^\eps(t,s,\nu)
 :=
 \mathbf{S}(t,s,\eps\nu),\qquad
 (t,s)\in[0,T]\times\R_+.
\end{equation*}
Then, the corresponding liquidity function is given by
\begin{equation*}
 \ell^\eps(t,s)
 :=
 \frac{1}{\eps}\ell(t,s),
 (t,s)\in[0,T]\times\R_+ \,.
\end{equation*}
Hence, as $\eps$ tends to zero, the market becomes completely
liquid.  So we expect the price of an option $V^\eps$ to converge to the
classical Black-Scholes price, $v^{BS}$, and we are interested in expansions
of the form
\begin{equation*}
V^\eps = v^{BS} + \eps v^{(1)} +\ldots + \eps^n v^{(n)}+ \ + o(\eps^{n}).
\end{equation*}

\noindent Indeed, we prove this type of results and identify the
functions $v^{(n)}$ in some cases.  In particular, we show that
\begin{equation}
\label{v-1}
v^{(1)}(t,s) =
\int_t^T \ \E_{t,s}
\left[
\frac{S_u^2\sigma^2(u,S_u)}{4\ell(u,S_u)}\left(v_{ss}^{BS}(u,S_u)\right)^2\right]du.
\end{equation}
This is exactly the liquidity premium of the standard Black-Scholes hedge.

\noindent The paper is organized as follows.
The problem is introduced in the next section and 
the approach is formally introduced in Section \ref{s.formal}.
Under a strong smoothness assumption,
full expansion is obtained in Section \ref{s.smooth}.
A quick convergence result is proved 
in Section \ref{s.conv}.  The Call option
is studied in Section \ref{s.call} and the Digital option in the final section.

\section{The general setting}
Let $(\Omega,\Fc,\P)$ be a complete probability space
endowed with a Brownian motion $W$ with completed
canonical filtration $\F=\{\Fc_t,t\in[0,T]\}$, where
$T>0$ is fixed maturity. The marginal price process $S_t$
is defined by the stochastic differential equation
\begin{equation*}
 \frac{dS_t}{S_t}=
 \sigma(t,S_t)dW_t,
\end{equation*}

\noindent where $\sigma$ is assumed to be bounded, Lipschitz-continuous and uniformly
elliptic.

\noindent Given a continuous portfolio strategy $Y$ with finite quadratic
variation process $\left<Y\right>$, the {\it small time
liquidation value of the portfolio} is given by
 \begin{equation*}
 dZ^{\eps, Y}_t
 =
 Y_t dS_t - \left[4\ell^\eps(t,S_t)\right]^{-1} d\langle Y\rangle_t
= Y_t dS_t - \eps\left[4\ell(t,S_t)\right]^{-1} d\langle Y\rangle_t.
 \end{equation*}
\noindent The dependence of the process $Z$ on its initial
condition is suppressed for simplicity.

\noindent Given a function $g:\R_+\longrightarrow\R$ satisfying
 \be\label{g}
 g &\mbox{is bounded from below and}&
 \sup_{s>0}\;\frac{g(s)}{1+s} \;<\; \infty ,
 \ee
the super-hedging cost is defined by
 \be\label{defV}
 V^\eps(t,s)
 &:=&
 \inf\left\{z\ :~ Z^{\eps, Y}_t=z \text{ and }Z^{\eps, Y}_T \ge g\left(S_T\right) 
                     ~\mbox{$\mathbb{P}$-$as$ for some}~Y\in\Ac_{t,s}
     \right\},
 \ee
\noindent where the time origin is removed
to $t$ and the initial condition for the price
process is $S_t=s$. We refer to \cite{cst}
for the precise definition of the set of admissible strategies $\Ac_{t,s}$.

\noindent This problem is similar to the super-replication
problem studied extensively in \cite{chsta,chstb,cstv,st00,st02,st1,st2}.
In the above setting, it is shown in \c{C}etin, Soner and Touzi
\cite{cst} that the value function of the super-hedging
problem is the unique viscosity solution of the 
following nonlinear equation, 
 \be \label{PDE}
 -V^\eps_t+\hat{H}^\eps\left(t,s,V^\eps_{ss}\right)
 \;=\;
 0,
 &\mbox{on}&
 [0,T)\times(0,\infty),
 \ee
\noindent satisfying the terminal condition
 $ V^\eps(T,.)=g$
\noindent and the growth condition
 \begin{equation}
 \label{e.growth}
 -C \;\le\; V^\eps(t,s) \;\le\; C(1+s),~(t,s)\in[0,T]\times\R_+,  \ \mbox{for some
 constant}\
 C>0\,.
 \end{equation}
\noindent Here, $\hat{H}^\eps$ denotes the elliptic majorant of the first guess operator $H^\eps$:
\begin{align*}
 \hat{H}^\eps(t,s,\gamma)&:=
 \sup_{\beta\ge 0}\;H^\eps(t,s,\gamma+\beta), \\
 H^\eps(t,s,\gamma)
 &:=
 -\frac12s^2\sigma^2(t,s)\gamma-\eps[4 \ell(t,s)]^{-1}s^2\sigma^2(t,s)\gamma^2.
 \end{align*}
\noindent By direct calculation, it follows that
\begin{equation*}
 \hat{H}^\eps(t,s,\gamma) =
 - \frac12 s^2\sigma^2(t,s)
 \left[\gamma+\left(\gamma + \frac{\ell(t,s)}{\epsilon}\right)^{-}+ \frac{\epsilon}{2\ell(t,s)}\left(\gamma+\left(\gamma + \frac{\ell(t,s)}{\epsilon}\right)^{-}\right)^2 \right].
\end{equation*}
\noindent For $\eps=0$, both $\hat{H}^\eps$, ${H}^\eps$ coincides with
the following standard elliptic operator,
\begin{equation*}
 \hat{H}^0(t,s,\gamma)
 =H^0(t,s,\gamma)
 =-\frac12s^2\sigma^2(t,s)\gamma,\qquad
 (t,s,\gamma)\in[0,T]\times\R_+\times\R .
\end{equation*}

\noindent Hence, the equation \reff{PDE} reduces to the linear Black-Scholes equation
 \be\label{BS}
  -\frac{\partial v^{BS}}{\partial t}-\frac12 s^2\sigma^2(t,s) v^{BS}_{ss}
   = 0.
\ee
\noindent We recall the well-known fact that its unique
 solution, $v^{BS}$, is the Black-Scholes price,
\begin{equation*}
 v^{BS}(t,s)
 =
 \E_{t,s}\left[g(S_T)\right],\qquad
 (t,s)\in[0,T]\times\R_+,
\end{equation*}
\noindent where we used the notation $\E_{t,s}=\E[\ \cdot \ |\ S_t=s]$.

\section{Formal calculations and Assumptions}
\label{s.formal}

\noindent It is formally clear that as the market
becomes more liquid, $V^\eps$ should converge
to the Black-Scholes price $v^{BS}$. Indeed, this
is proved in Section \ref{s.conv}.  We are also interested
in a Taylor expansion of $V^\eps$ in the parameter $\eps$, i.e.,
\begin{equation}
\label{Taylor}
V^\eps(t,s) = v^{BS}(t,s)
+ \eps v^{(1)}(t,s) + \eps^{(2)} v^2(t,s) + \ldots + \eps^n v^{(n)}(t,s) + o(\eps^n),
\end{equation}
\noindent where $o(\eps^n)$ is the standard notation, indicating that
$o(\eps^n)/\eps^n$ converges to zero as $\eps$ tends to zero.

\noindent Indeed, under sufficient regularity
\begin{equation*}
v^{(n)}(t,s) =\frac1{n !}\  \left. \frac{\partial^n V^\eps(t,s)}{\partial \eps^n}\right|_{\eps=0}.
\end{equation*}

\noindent Thus, formally differentiate the equation \reff{PDE} $n$-times
with respect to $\eps$
and then set $\eps$ to zero.  Using the above
formal definition of $v^{(n)}$, we arrive at,
\begin{eqnarray}
\label{e-n}
0&=&-v^{(n)}_t - \frac12 s^2 \sigma^2(t,s) v^{(n)}_{ss}-
F_n(t,s),\\
\label{f-n}
F_n(t,s)&=& \frac{s^2\sigma^2(t,s)}{4\ell(t,s)}\sum_{k=0}^{n-1}
\left[ v_{ss}^{(k)}(t,s)\ v_{ss}^{(n-1-k)}(t,s)\right] ,
\end{eqnarray}
\noindent where we set  $v^{(0)}:=v^{BS}$.  For all $n\ge 1$,
the terminal data is $v^{(n)}(T,\cdot)\equiv 0$, so that
the Feymann-Kac formula yields
\begin{equation}
\label{v-n}
v^{(n)}(t,s) = \sum_{k=0}^{n-1} \  \E_{t,s}
\left[\int_t^T \left(\frac{S_u^2\sigma^2}{4\ell}v_{ss}^{(k)}
v_{ss}^{(n-1-k)}\right)(u,S_u)du\right].
\end{equation}
\noindent In particular, $v^{(1)}$ is given as in \reff{v-1}.

\noindent The above calculations yield a rigorous
proof when the pay-off is sufficiently regular.
We will prove this in Section \ref{s.smooth}.
On the other hand,
for some discontinuous pay-offs the above functions
may not be finite. For instance, for a digital option, $v^{(1)} \equiv \infty$. Indeed, if we take
$$
g(s):=\mathds{1}_{s\geq K},\qquad \sigma(t,s)\equiv \sigma \qquad{\mbox and}
\qquad \ell(t,s)\equiv \ell,
$$

\noindent we compute that 

\begin{align*}
v^{(1)}(t,s)&=\frac{1}{8\pi\ell\sigma^2}\int_t^T \frac{(u-t)e^{-\left(\frac{1}{\sigma \sqrt{T+u-2t}} \ln\left(\frac sK\right) + \frac{\sigma}{2} \frac{T-2u+t}{\sqrt{T+u-2t}}\right)^2}}{(T-u)^{\frac32}(T+u-2t)^{\frac32}}\\
&+\frac{1}{8\pi\ell\sigma^2}\int_t^T \frac{e^{-\left(\frac{1}{\sigma \sqrt{T+u-2t}} \ln(\frac sK) + \frac{\sigma}{2} \frac{T-2u+t}{\sqrt{T+u-2t}}\right)^2}}{\sqrt{T-u}(T+u-2t)^{\frac32}}\left(\frac{\ln\left(\frac sK\right)}{\sigma \sqrt{T+u-2t}}  + \frac{\sigma}{2} \frac{T-2u+t}{\sqrt{T+u-2t}}\right)^2.
\end{align*}

\noindent The first term above is actually $+\infty$ because of the non-integrability of $(T-u)^{-3/2}$ near $T$.

\vspace{0.5em}
\noindent In such cases, the expansion is not valid and a careful study of the
behavior of $V^\eps$ near the terminal data is needed. This will be done in Section \ref{s.digital}. However,
we first prove the full expansion in the "smooth" case.  Then,
 in Section \ref{s.call}, we consider the Call option proving the
expansion up to $n=2$.
Clearly, this later result extends to all Put options.  Also, remarks on other
payoffs and higher expansions are given in Remarks \ref{r.bs} and \ref{r.higher}.

\section{Expansion for smooth pay-offs}
\label{s.smooth}

\noindent In this section, we prove the expansion under the
assumption
that  there is a constant $\hat C$ so that
\begin{eqnarray}
\label{v.smooth}
-\hat{C} \le v^{(n)}(t,s) \le \hat{C}(1+s), & &
\left|(s^2+1)v^{(n)}_{ss}(t,s)\right|\le \hat{C},\\
\nonumber
\left|F_n(t,s)\right| \le \hat{C}, & & \qquad
\forall \ (t,s) \in [0,T]\times \R_+, \ n=1,2,\ldots.
\end{eqnarray}
\noindent Clearly, this is an implicit assumption on the
pay-off $g$.  Essentially, it holds for all smooth
pay-offs growing at most linearly.  In particular,
\reff{v.smooth} holds if $\sigma(t,s)\equiv \sigma$,
$\ell(t,s)\equiv \ell$ and if there exists a constant
$C$ so that

$$
-C \le g(s) \le C(1+s), \ \ \left|(s^2+1)
\frac{\partial^n}{\partial s^n} g(s)\right| \le C, \qquad
\forall \ s \in \R_+,\ n=2,3,\ldots.
$$
\noindent This is proved by using the homogenity of the Black-Scholes equation
and differentiating it repeatedly.

\noindent Following the techniques developed
in the papers \cite{fso,fs89,lsst,sszj,son93},
for an integer $n\ge 0$ we define,
\begin{equation}
\label{v.n}
V^{\eps,n}(t,s):=\frac{V^\eps(t,s)-\sum_{k=0}^{n-1}\eps^k
 v^{(k)}(t,s)}{\eps^{n}}\ ,
\end{equation}
\noindent where as before we set $v^{(0)}=v^{BS}$.
\begin{Theorem}
\label{t.conv-smooth}  Assume \reff{v.smooth}.
\noindent Then, for every $n=1,2,\ldots$, there are constants $C_n$ and $\eps_0>0$
so that for every $\eps
\in(0,\eps_0]$, and  $n=1,2,\ldots$,
\begin{equation}
\label{smooth-est}
v^{BS}(t,s) \le V^\eps(t,s) \le v^{\eps,n}(t,s):=
\sum_{k=0}^{n-1}\ [\eps^k v^{(k)}(t,s)] +\eps^n {C_n}(T-t).
\end{equation}
\noindent In particular,  as $\eps \downarrow 0$, $V^\eps$
converges to the
Black-Scholes price $v^{BS}$ uniformly on compact sets.
Moreover, for every $n\ge 1$, $V^{\eps,n}$ converges to
$v^{(n)}$, again  uniformly on compact sets.
\end{Theorem}

\proof
Clearly, $v^{BS} \le V^\eps$.
\noindent We continue by proving the upper bound. Let $v^{\eps,n}$
be as in \reff{smooth-est} with a constant $C_n$ to be
determined below.
Using \reff{e-n}, we calculate that
\begin{eqnarray*}
-v^{\eps,n}_t(t,s) & + & \hat H^\eps(t,s,v^{\eps,n}_{ss}(t,s)) \ge
-v^{\eps,n}_t(t,s)  +  H^\eps(t,s,v^{\eps,n}_{ss}(t,s)) \\
& = &  -v^{\eps,n}_t -\frac12 s^2 \sigma^2 v^{\eps,n}_{ss}
- \frac{\eps s^2 \sigma^2}{4\ell(t,s)}
\left(v^{\eps,n}_{ss}\right)^2\\
&= & \eps^n{C_n} +\sum_{k=1}^{n-1}\ [ \eps^k\ F_k(t,s)]
- \frac{\eps s^2 \sigma^2}{4\ell(t,s)}
\left(v^{\eps,n}_{ss}\right)^2.
\end{eqnarray*}
In view of \reff{f-n},
$$
\frac{\eps s^2 \sigma^2}{4\ell(t,s)}
\left(v^{\eps,n}_{ss}\right)^2-\sum_{k=1}^{n-1}\ [ \eps^k\ F_k(t,s)]
 = \eps^n F_n(t,s) +
\eps^{n+1} \frac{s^2 \sigma^2}{4\ell(t,s)}\ g^\eps(t,s),
$$
where $g^\eps(t,s)$ is a quadratic function
$v^{(k)}_{ss}(t,s)$ for $k\le n$ and possibly powers of
$\eps$.  Hence by
\reff{v.smooth}, there is a constant $C_n$,
$$
\left|\sum_{k=1}^{n-1}\ [ \eps^k\ F_k(t,s)]
- \frac{\eps s^2 \sigma^2}{4\ell(t,s)}
\left(v^{\eps,n}_{ss}\right)^2 \right|
\le  \eps^n C_n.
$$
Hence, we conclude that $v^{\eps,n}$ is a supersolution
of \reff{PDE}.  Moreover, by \reff{v.smooth},
$-C \le v^{\eps,n}(t,s) \le C(1+s)$.
Then, by the comparison theorem for \reff{PDE} (Theorem 6.1 of
\cite{cst}), we conclude that
$V^\eps(t,s) \le v^{\eps,n}(t,s)$.

\noindent In particular, this estimate implies the
convergence of $V^\eps$ to $v^{BS}$.
To prove the convergence of $V^{\eps,n}$,
we first observe that
$$
V^\eps=
\sum_{k=0}^n\ [\eps^k v^{(n)}(t,s)]+\eps^n V^{\eps,n}.
$$
\noindent Using the equations \reff{PDE} and \reff{e-n}, we conclude that
$V^{\eps,n}$ is a viscosity solution of
 \b*
 -V^{\eps,n}_t - \frac12 s^2\sigma^2(t,s) V^{\eps,n}_{ss}
 +F^{\eps,n}\left(t,s,V^{\eps,n}_{ss}\right)=0,
 &&
 (t,s)\in[0,T)\times \R_+,
 \e*
where
 \b*
 F^{\eps,n}(t,s,\gamma)
 :=
 \frac1{\eps^n}\ \left[ \hat H^\eps(t,s,v^{\eps,n}_{ss}(t,s)+\eps^n \gamma)
 + \frac12 s^2 \sigma^2 v^{\eps,n}_{ss}
 + \sum_{k=1}^{n-1} \ \eps^k F^k(t,s) \right].
 \e*
Tedious but a straightforward calculation shows that
 \b*
 \lim_{(t',s',\gamma',\eps)\to (t,s,\gamma,0)} F^{\eps,n}(t',s',\gamma')
 &=&
 F^{n}(t,s),
 \e*
 where $F_n$ is as in \reff{f-n}.
\noindent Then, by the classical stability results of viscosity solutions
\cite{bp,cil,fs}, the Barles-Perthame semi-relaxed limits
 \b*
 \underline{v}^{(n)}(t,s):=\liminf_{(t',s',\eps)\to(t,s,0)} V^{\eps,n}(t',s')
 ~\mbox{and}~
 \overline{v}^{(n)}(t,s):=\limsup_{(t',s',\eps)\to(t,s,0)} V^{\eps,n}(t',s'),
 & &
 \e*
\noindent are, respectively, a viscosity supersolution and a subsolution of the equation
\reff{e-n} satisfied by $v^{(n)}$. Moreover it follows from \reff{smooth-est} that

$$
 \underline{v}^{(n)}(T,\cdot)= \overline{v}^{(n)}(T,\cdot)=0=v^{(n)}(T,\cdot).
$$
\noindent We now use the comparison result for the linear partial differential equation \reff{e-n},
and conclude that $\underline{v}^{(n)} \ge \overline{v}^{(n)}$.
Since
$$
\underline{v}^{(n)}(t,s) \le \liminf_{\eps\to 0}V^{\eps,n}(t,s)
\le \limsup_{\eps\to 0}V^{\eps,n}(t,s)\le \overline{v}^{(n)}(t,s)
$$
on $[0,T]\times \R_+$,
this proves that $\underline{v}^{(n)}= \overline{v}^{(n)}=v^{(n)}$.
Hence, $V^{\eps,n}$ converges to the unique
solution $v^{(n)}$, uniformly on
compact sets.

\ep

\section{A general convergence result}
\label{s.conv}

\noindent In this section, we prove an easy convergence result
under the following general assumption.  We assume that
\begin{equation}
c s^2\le \ell(t,s),
\label{eqq}
\end{equation}

\noindent for some constant and 

\begin{Assumption}\label{assump.smooth}
There is a decreasing sequence of smooth approximation
$\ g_m\ge g$ of the pay-off $g$ satisfying \reff{v.smooth}
with $n=1,2$. Let $v^{(n)}_m$, $F^{n}_m$ be the previously
defined functions with pay-off $g_m$.
Then, $F^1_m(t,s) \le c_m$ for some constant $c_m$.
\end{Assumption}

\noindent This assumption is satisfied by all Lipschitz
or for all bounded pay-offs.

\begin{Theorem}
\label{t.conv}  Assume \reff{g}, \reff{eqq} and that Assumption \ref{assump.smooth} holds true.
Then, as the liquidity parameter goes to infinity,
or equivalently as $\eps \downarrow 0$, $V^\eps$
converges to the
Black-Scholes price $v^{BS}$.
\end{Theorem}

\proof  Let $c_m$ be as above and set
$$
u^\eps(t,s):= v^{BS}_m(t,s) + \eps c_m (T-t).
$$
As in the proof of Theorem \ref{t.conv-smooth},
we can show that $u^\eps$ is a super-solution
of \reff{PDE}.  Hence, $V^\eps \le u^\eps$.
Therefore,
$$
\limsup_{\eps \downarrow 0} \ V^\eps(t,s)
 \le v^{BS}_m(t,s).
 $$
 By \reff{g}, $v^{BS}_m(t,s)$ converges to
$v^{BS}(t,s)$.  Since $V^\eps \ge v^{BS}$,
this proves the convergence of $V^\eps$
to $v^{BS}$.

\ep

\section{First order expansion for convex payoffs}
\label{s.call}

\noindent One major limitation of our previous result is that the Call pay-off does not satisfy the Assumption  \reff{v.smooth}. Therefore, in this section, we prove the first term
in the Taylor expansion \reff{Taylor}, i.e.,
\begin{equation}
\label{expansion}
V^\eps(t,s) = v^{BS}(t,s)
+ \eps v^{(1)}(t,s)+ o(\eps),
\end{equation}
for convex payoffs satisfying weaker assumptions than \reff{v.smooth}. In particular, we will show that call options verify those assumptions. 

\subsection{The general result}

\noindent In order to capitalize on the results we have already obtained for smooth payoffs, we will also consider a regularized version of our problem

\begin{align}
\label{e.approxcall}
\nonumber -&V^{\epsilon,\alpha}_t+ \widehat{H}^{\eps}(t,s,V^{\epsilon,\alpha}_{ss})=0, \text{  for $(t,s)\in [0,T)\times \mathbb{R}_+$},\\
& V^{\eps,\alpha}(T,s)=\widehat{g}_{\alpha}(s),
\end{align}
\noindent where $\widehat{g}_{\alpha}(s)=\phi_{\alpha}\ast g(s)$ with $\phi_{\alpha}(\cdot):=\frac{1}{\alpha}\phi(\frac{\cdot}{\alpha})$ and $\phi$ is a positive, symmetric bump function on $\mathbb{R}$, compactly supported in $[-1,1]$ and satisfying 

$$
\int_{-1}^1{\phi(u)du}=1.
$$

\noindent By convexity of $g$, for all $\alpha >0$ we have $\widehat{g}_{\alpha} \geq g$, so that by monotony of our problem

$$
V^{\eps}\leq V^{\eps,\alpha}.
$$

\noindent Thus, since the main idea of our proof is to find a super-solution of (\ref{PDE}), we see that it is enough to find a super-solution of (\ref{e.approxcall}). Let $v^{BS,\alpha}$ and $v^{(1),\alpha}$, respectively, be the Black-Scholes price and the first-order expansion term for the regularized option. We now state our assumptions

\begin{Assumption} \label{hyp.call}
\begin{itemize}
\item[\rm{(i)}] $v^{BS}+v^{BS,\alpha}+v^{(1)}+v^{(1),\alpha}<+\infty$.
\item[\rm{(ii)}] As $\alpha$ tends to $0$ we have
\begin{align*}
v^{BS,\alpha}(t,s) &=v^{BS}(t,s) +O(\alpha^2),\\
v^{(1),\alpha}(t,s)&=v^{(1)}(t,s)+o(1).
\end{align*}
\item[\rm{(iii)}] There exists a constant $c_*$ independent of $s$, $T-t$ and $\alpha$ and $(\nu,\beta)\in [0,1]\times[1/2,1]$ such that $1<2\beta+\nu<2$ and
$$
\frac{s^2\sigma^2}{4\ell} (v^{(1),\alpha}_{ss}(t,s))^2 \le \frac{c_*}{(T-t)^{1-\nu}\alpha^{2+2\nu}} \text{,}\qquad s\left|v^{BS,\alpha}_{ss}(t,s)\right| \le \frac{c_*}{(T-t)^{1-\beta}\alpha^{2\beta-1}}.
$$
\end{itemize}
\end{Assumption}

\noindent This assumption will be proved to be verified by Call options payoffs in subsection \ref{exp.call}.

\vspace{0.8em}
\noindent Let $V^{\epsilon,1}$ be as (\ref{v.n}), i.e.
\begin{equation*}
V^{\epsilon,1}(t,s):=\frac{V^{\epsilon}(t,s)-v^{BS}(t,s)}{\epsilon}.
\end{equation*}
\begin{Theorem}
\label{t.call}
Let Assumption \ref{hyp.call} hold true and let $a \in (\frac12,\frac{1}{2\beta+\nu})$. Then for every $(t,s)\in[0,T]\times\R_+$ we have,
\begin{equation*}
v^{BS}\le V^{\eps} \le v^{BS,\epsilon^a} + \epsilon v^{(1),\epsilon^a} + c_*(T-t)^{\beta+ \frac{\nu-1}{2}}\epsilon^{2-a(\nu+2\beta)} + c_*(T-t)^\nu\epsilon^{3-2a(1+\nu)}.
\end{equation*}

\noindent Moreover, $V^{\eps}\rightarrow v^{BS}$, $V^{\eps,1}\rightarrow v^{(1)}$ uniformly on compact sets, and \reff{expansion} holds true.
\end{Theorem}

\proof It is clear that $V^\eps \ge v^{BS}$.  To prove the
reverse inequality, we start by following a technique similar to the one used in the proof
of Theorem \ref{t.conv-smooth}. Set
$$
v^{\eps,2}:= v^{BS,\epsilon^a} + \epsilon v^{(1),\epsilon^a} + c_*(T-t)^{\beta+ \frac{\nu-1}{2}}\epsilon^{2-a(\nu+2\beta)} + c_*(T-t)^\nu\epsilon^{3-2a(1+\nu)}.$$
\noindent We calculate that for $(t,s) \in [0,T)\times \R_+$
\begin{align*}
&-v^{\eps,2}_t  +  \hat H^\eps(t,s,v^{\eps,2}_{ss}) \ge
-v^{\eps,2}_t +  H^\eps(t,s,v^{\eps,2}_{ss}) \\
& =  \frac{c_*\epsilon^{2-a(\nu+2\beta)}}{(T-t)^{1-\beta- \frac{\nu-1}{2}}} +\frac{c_*\epsilon^{3-2a(1+\nu)}}{(T-t)^{1-\nu}} -v^{BS,\epsilon^a}_t
- \eps v^{(1),\epsilon^a}_t -\frac{1}{2}
s^2 \sigma^2 v^{\eps,2}_{ss}
-\frac{ \eps s^2 \sigma^2}{4\ell}
\left(v^{\eps,2}_{ss}\right)^2\\
&=  \frac{c_*\epsilon^{2-a(\nu+2\beta)}}{(T-t)^{1-\beta- \frac{\nu-1}{2}}} +\frac{c_*\epsilon^{3-2a(1+\nu)}}{(T-t)^{1-\nu}}  - \frac{s^2\sigma^2}{4\ell} (v^{(1),\epsilon^a}_{ss})^2\eps^3
- \frac{s^2\sigma^2}{2\ell}v^{BS,\epsilon^a}_{ss} v^{(1),\epsilon^a}_{ss}\eps^2.
\end{align*}
\noindent In view of Assumption \ref{hyp.call}$\rm{(iii)}$, this quantity is always positive. 
We now analyze the terminal condition. In view of the conditions imposed on $a,\beta$ and $\nu$
\begin{equation*}
v^{\eps,2}(T,s) = v^{BS,\epsilon^a}(T,s) = \widehat{g}_{\eps^a}(s).
\end{equation*}

\noindent Hence, $v^{\eps,2}$ is a super-solution of \reff{e.approxcall} and therefore of \reff{PDE}.
Then, by the comparison theorem for \reff{PDE} (proved in
\cite{cst}), we conclude that
$V^\eps(t,s) \le v^{\eps,2}(t,s)$.

\vspace{0.6em}
\noindent We now let $\epsilon$ go to $0$ in the above inequalities. This proves that $V^{\eps}$ converges to $v^{BS}$ uniformly on compact sets.

\vspace{0.6em}
\noindent Finally, by Assumption \ref{hyp.call}$\rm{(ii)}$ 
\begin{equation*}
0\leq V^{\eps,1}(t,s)\leq v^{(1)}(t,s) + o\left(\eps^{\min\left\{1-a(2\beta+\nu),2-2a(1+\nu)\right\}}\right) +O\left(\eps^{2a-1}\right),
\end{equation*}
\noindent where it is clear with our conditions on $a,\beta$ and $\nu$ that the $o(\cdot)$ and  $O(\cdot)$ above go to $0$ as $\eps$ tends to $0$.

\vspace{0.6em}
\noindent Using this estimate, we then prove the convergence of $V^{\eps,1}$
exactly as in Theorem \ref{t.conv-smooth}.
\ep

\begin{Remark}
\label{r.higher}
{\rm{Higher expansions can be proved similarly, provided that we extend Assumption \ref{hyp.call} for $n\ge 2$.}}
\end{Remark}

\subsection{Expansion for the Call option}\label{exp.call}

In this section, we take
$$
g(s)=(s-K)^+, \qquad \sigma(t,s)\equiv \sigma,
\qquad \ell(t,s)\equiv \ell,
$$
\noindent and we verify that Assumptions \ref{hyp.call}$\rm{(ii)}$ and \ref{hyp.call}$\rm{(iii)}$ are satisfied, since Assumption \ref{hyp.call}$\rm{(i)}$ is trivial. 

\vspace{0.5em}
\noindent Straightforward but tedious calculations using the Feynman-Kac formula yield
\begin{eqnarray*}
\label{e.call}
v^{BS,\alpha}_{ss}(t,s)&=&\frac{1}{\sigma s\sqrt{2\pi \tau\ }}\int_{-1}^{1}\phi(u)\exp{\left(- \frac{1}{2} d_1(s,K+\alpha u,\tau)^2\right)}du,\\[0.8em]
\nonumber
v^{(1),\alpha}(t,s)&=&\frac{1}{8\ell \pi }\int_0^{\tau}\int_{-1}^1\int_{-1}^1\frac{\phi(x)\phi(y)h_{\alpha}(\tau,v,s,K,x,y)}{ \sqrt{ v(2\tau-v)}}dxdydv,
\end{eqnarray*}
where
\begin{align*}
\tau&=T-t,\\[0.5em]
d_1(s,k,t)  &= \frac{1}{\sigma \sqrt{t}} \ln(s/k)+ \frac{1}{2}\sigma \sqrt{t},\\[0.8em]
\delta(\tau,v,s,k) &= \frac{1}{\sigma \sqrt{2\tau-v}} \ln(s/k) - \frac{\sigma}{2} \frac{\tau-2v}{\sqrt{2\tau-v}},\\[0.8em]
h_{\alpha}(\tau,v,s,k,x,y)&=\exp\left(-\delta(\tau,v,s,k)^2+ \frac{\delta(\tau,v,s,k)}{\sigma\sqrt{2\tau-v}}\left(\log\left(1+ \frac{\alpha x}{k}\right)+\log\left(1+ \frac{\alpha y}{k}\right)\right)\right)\\
&\times \exp\left(- \frac{\tau}{2\sigma^2v(2\tau-v)}\left(\log\left(1+ \frac{\alpha x}{k}\right)-\log\left(1+ \frac{\alpha y}{k}\right)\right)^2\right)\\
&\times \exp\left(- \frac{1}{\sigma^2(2\tau-v)}\log\left(1+ \frac{\alpha x}{k}\right)\log\left(1+ \frac{\alpha y}{k}\right)\right).
\end{align*}

\noindent The following two propositions, whose proof is relagated to the appendix, ensure that Assumptions \ref{hyp.call}$\rm{(ii)}$ and \ref{hyp.call}$\rm{(iii)}$ are satisfied 

\begin{Proposition} \label{prop.ineq}
There exists a constant $c_*$, independent of $s$, $\tau$ and $\alpha$ so that for all $(\nu,\beta)\in [0,1]\times[1/2,1]$:
\begin{equation*}
s\left|v^{BS,\alpha}_{ss}(t,s)\right| \le \frac{c_*}{\tau^{1-\beta}\alpha^{2\beta-1}} \text{, }\qquad \frac{s^2\sigma^2}{4\ell} (v^{(1),\alpha}_{ss}(t,s))^2 \le \frac{c_*}{\tau^{1-\nu}\alpha^{2+2\nu}}.
\end{equation*}
\end{Proposition}

\begin{Proposition}\label{prop.exp}
\noindent As $\alpha$ tends to $0$ we have the following expansions

\begin{align*}
v^{BS,\alpha}(t,s)&=v^{BS}(t,s)+\alpha^2 \frac{e^{- \frac{1}{2}d_0(s,K,\tau)^2}}{2K\sigma\sqrt{2\pi\tau}}\int_{-1}^1\phi(v)v^2dv+O(\alpha^4),\\[0.8em]
v^{(1),\alpha}(t,s)&=v^{(1)}(t,s)- \alpha\frac{e^{- \frac{1}{2}d_0(s,K,\tau)^2}}{8K\sigma\ell\sqrt{2\pi\tau}}\int_{-1}^1\int_{-1}^1\phi(x)\phi(y)|x-y|dxdy+o(\alpha),
\end{align*}

\noindent where $d_0(s,k,\tau)=\frac{1}{\sigma \sqrt{\tau}} \ln(s/k)- \frac{1}{2}\sigma \sqrt{\tau}$.
\end{Proposition}

\begin{Remark}
\label{r.bs}
{\rm{ It is not hard to show that the results of Propositions \ref{prop.ineq} and \ref{prop.exp} hold for all convex linear combination of call or put options. However, we cannot use the above proof for, say, a call spread option whose payoff is neither convex nor concave.}}
\end{Remark}

\subsection{Numerical Experiments}\label{call.num}
In order to have a better grasp of the liquidity effects, we also solved numerically (with simple finite difference methods) the PDE \reff{PDE}. We represent below the behaviour of the liquidity premium (that is to say $V^\eps-v^{BS}$) when the time to maturity $t$ and the spot price vary

\vspace{1.2em}
\begin{figure}[!ht]
\begin{center}

   \includegraphics[height=24em]{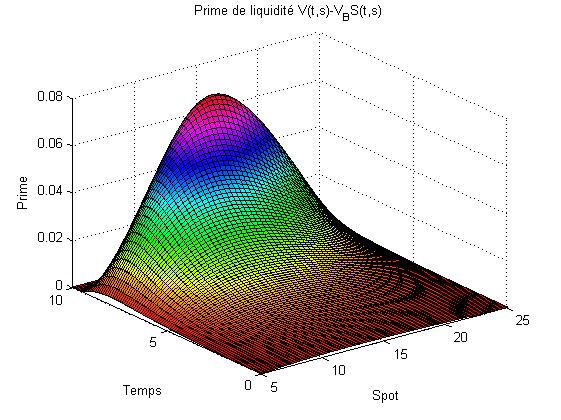}
   \caption{Call liquidity premium - $T=10$, $K=15$, $\sigma=0.5$, $\epsilon=0.1$, $\ell=1$}
   \end{center}

\end{figure}

\noindent In the above figure, the liquidity effect is strongly marked for ATM options and disapears quickly for ITM and OTM options. This was to be expected. Indeed, our calculations showed that the liquidity effect is, for the first order, driven by the $\Gamma$ of the call option (see \reff{gamma.call}), which explodes for ATM options near maturity. Moreover, with our set of parameters, the first order correction is at most $0.06$ for a BS price of $8.56$, which means that the hedge against liquidity risk is not that expensive when the illiquidity is not too strong.

\vspace{2.8em}
\noindent We now compare the real liquidity premium with its first-order expansion term.

\begin{figure}[!ht]
\begin{center}

   \includegraphics[height=25em]{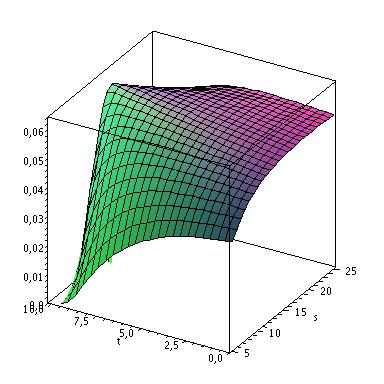}
   \caption{Call first order liquidity premium - $T=10$, $K=15$, $\sigma=0.5$, $\epsilon=0.1$, $\ell=1$}
   \end{center}

\end{figure}

\vspace{1.2em}
\noindent A rapid examination of the above figure shows that the first order approximation remains excellent as long as we do not go too far from the maturity time $T$ and we stay close to the money $s=K$. Otherwise, the first order overvalues the liquidity premium.
\section{Digital Option} \label{s.digital}

In this section, we analyze the specific example of a Digital option in the context of Black-Scholes model with constant liquidity parameter
$$
g(s):=\mathds{1}_{s\geq K},\qquad {\mbox and}\qquad \sigma(t,s)\equiv \sigma,
\qquad \ell(t,s)\equiv \ell.
$$
\subsection{Theoretical bounds}
As pointed out earlier, for the Digital option, the first-order term that we obtained formally is equal to $+\infty$. Thus, the expansion (\ref{Taylor}) is no longer valid and our aim in this section is to find bounds for the first-order of the expansion. We start by approximating the option by a sequence of regularized call spreads. Then the original problem \reff{PDE} is replaced by

\begin{align}
\label{e.approxdig}
\nonumber -&V^{\epsilon,\alpha}_t+ \widehat{H}^{\eps}(t,s,V^{\epsilon,\alpha}_{ss})=0, \text{  for $(t,s)\in [0,T)\times \mathbb{R}_+$,}\\
& V^{\eps,\alpha}(T,s)=\widehat{g}_{\alpha}(s),
\end{align}
where $\widehat{g}_{\alpha}(s)=\phi_{\alpha}\ast g_{\alpha}(s)$ with $g_{\alpha}(s)=\frac{(s-K+2\alpha)^+-(s-K+\alpha)^+}{\alpha}$.

\vspace{0.5em}
\noindent Since $\phi_\alpha$ has compact support in $[-\alpha,\alpha]$, notice that $\widehat g_\alpha \geq g$. Then, since the terminal condition is smooth, it follows from the comparison principle that

\begin{equation}
\label{e.vv}
V^{\eps}(t,s)\leq V^{\epsilon,\alpha}(t,s),\text{  for $(t,s,\alpha)\in [0,T]\times \mathbb{R}_+\times \mathbb{R}^{*}_{+}$}.
\end{equation}

\vspace{0.5em}
\noindent With the same notations as in the previous section, we directly calculate using again the Feynman-Kac formula that

\begin{eqnarray*}
\label{e.dig}
v^{BS,\alpha}_{ss}(t,s)&=&\frac{1}{\sigma s\alpha\sqrt{2\pi \tau\ }}\int_{-1}^{1}\phi(u)\left(e^{- \frac{1}{2} d_1(s,K+\alpha u-2\alpha,\tau)^2}-e^{- \frac{1}{2} d_1(s,K+\alpha u-\alpha,\tau)^2}\right)du,\\[0.8em]
\nonumber
v^{(1),\alpha}(t,s)&=&\frac{1}{8\ell \pi \alpha^2}\int_0^{\tau}\int_{-1}^1\int_{-1}^1\frac{\phi(x)\phi(y)\widehat{h}_{\alpha}(\tau,v,s,K,x,y)}{ \sqrt{ v(2\tau-v)}}dxdydv,
\end{eqnarray*}
where

\begin{align*}
\widehat{h}_{\alpha}(\tau,v,s,K,x,y)&=\sum_{1\leq i,j\leq 2}h_{\alpha}(\tau,v,s,K,x-i,y-j).
\end{align*}

\vspace{1em}
\noindent Then, we have the two following propositions which are proved exactly as in the call option case (since the functions involved here are essentially the same)

\begin{Proposition} \label{prop.ineqdig}
There exists a constant $c_*$, independent of $s$, $\tau$ and $\alpha$ so that for all $(\nu,\beta)\in [0,1]\times[1/2,1]$
\begin{align*}
\nonumber s\left|v^{BS,\alpha}_{ss}(t,s)\right| \le \frac{c_*}{\tau^{1-\beta}\alpha^{2\beta}} \text{, } \qquad \frac{s^2\sigma^2}{4\ell} (v^{(1),\alpha}_{ss}(t,s))^2 \le \frac{c_*}{\tau^{1-\nu}\alpha^{6+2\nu}}.
\end{align*}
\end{Proposition}

\vspace{0.5em}
\begin{Proposition}\label{prop.expdig}
\noindent As $\alpha$ tends to $0$ we have the following expansions:

\begin{align*}
v^{BS,\alpha}(t,s)&=v^{BS}(t,s)+ \frac32\alpha \frac{e^{- \frac{1}{2}d_0(s,K,\tau)^2}}{K\sigma\sqrt{2\pi\tau}}+O(\alpha^2),\\[0.8em]
v^{(1),\alpha}(t,s)&=\alpha^{-1}\frac{e^{- \frac{1}{2}d_0(s,K,\tau)^2}}{8K\sigma\ell\sqrt{2\pi\tau}}\int_{-1}^1\int_{-1}^1\scriptstyle\phi(x)\phi(y)(|x-y-1|+|x-y+1|-2|x-y|)dxdy \displaystyle+o(\alpha^{-1}).
\end{align*}
\end{Proposition}

\vspace{1em}
Define $V^{\epsilon,1,c}$ by
\begin{equation*}
V^{\epsilon,1,c}(t,s):=\frac{V^{\epsilon}(t,s)-v^{BS}(t,s)}{\epsilon^{c}}.
\end{equation*}

\vspace{2em}

\begin{Theorem}
\label{t.digital}
Let $(\beta,\nu) \in [1/2,1]\times[0,1]$ be such that $\gamma :=\frac{2\beta
+\nu-1}{2\beta+\nu+4}\in (0,1)$ and set $a:=\frac25(1-\gamma)$. Then for all $(t,s)\in[0,T]\times\R_+$,

\begin{equation*}
v^{BS}\le V^{\eps} \le v^{BS,\epsilon^a} + \epsilon v^{(1),\epsilon^a} + c_*(T-t)^{\beta+ \frac{\nu-1}{2}}\epsilon^{2-3a-a(\nu+2\beta)} + c_*(T-t)^\nu\epsilon^{3-2a(3+\nu)}.
\end{equation*}
\noindent In particular, $V^{\eps}$ converges to $v^{BS}$,  uniformly on compact sets and

\begin{equation*}
0 \leq \liminf_{(t',s',\eps)\to(t,s,0)} V^{\eps,1,a}(t',s',a)\leq \limsup_{(t',s',\eps)\to(t,s,0)} V^{\eps,1,a}(t',s')\leq \frac32 \frac{e^{- \frac{1}{2}d_0(s,K,\tau)^2}}{K\sigma\sqrt{2\pi\tau}} + c_*(T-t)^{\frac{5\gamma}{2(1-\gamma)}},
\end{equation*}
$i.e.$ the order of the expansion is at least $2/5$.

\end{Theorem}
\proof It is clear that $V^\eps \ge v^{BS}$.  To prove the
reverse inequality, we start by following a technique similar to the one used in the proof
of Theorem \ref{t.call}. Set
\begin{equation*}
v^{\eps,2}:= v^{BS,\epsilon^a} + \epsilon v^{(1),\epsilon^a} + c_*(T-t)^{\beta+ \frac{\nu-1}{2}}\epsilon^{2-3a-a(\nu+2\beta)} + c_*(T-t)^\nu\epsilon^{3-2a(3+\nu)}.
\end{equation*}
We proceed exactly as in Theorem \ref{t.call} using Proposition \ref{prop.ineqdig}. The result is 

\begin{equation*}
-v^{\eps,2}_t(t,s)  +  \hat H^\eps(t,s,v^{\eps,2}_{ss}(t,s)) \ge 0 \text{, for $(t,s) \in [0,T)\times \R_+$}.
\end{equation*}
We now analyze the terminal condition. Since $2\beta +\nu >1$, we have
\begin{equation*}
v^{\eps,2}(T,s) = v^{BS,\epsilon^a}(T,s).
\end{equation*}

\noindent Hence, $v^{\eps,2}$ is a super-solution of \reff{e.approxcall} and therefore of \reff{PDE}.
Then, by the comparison theorem for \reff{PDE} (proved in
\cite{cst}), we conclude that
$V^\eps(t,s) \le v^{\eps,2}(t,s)$.

\noindent Then by Proposition \ref{prop.expdig} and the conditions imposed on $a$, $\beta$ and $\nu$, we obtain easily the uniform convergence on compact sets of $V^\eps$ to $v^{BS}$ by letting $\eps$ go to $0$. 

\noindent Now for the first order term, we would like to use our expansions and obtain a finite majorant for $V^{\eps,1,c}$ with the largest possible $c$. It is easy to argue that $c=a$ is the best choice possible. This, in turn, imposes the following condition

\begin{equation*}
a\leq \min\left\{\frac12,\frac{2}{4+2\beta + \nu},\frac{3}{7+2\nu}\right\}=\frac{2}{4+2\beta + \nu}.
\end{equation*}

\noindent Now it follows that, for all $\gamma>0$ small enough, there are $\beta$ and $\nu$ satisfying our conditions so that $\frac{2}{4+2\beta + \nu}=\frac25(1 -\gamma)$. It suffices then to take the $\liminf$ and $\limsup$ in the inequality to prove the result.
\ep

\subsection{Numerical results}

\paragraph{The digital option liquidity premium}

In this section, we provide numerical results for the case of the Digital option. As in the section \ref{call.num} the PDE \reff{PDE} is solved with finite difference method. We represent below the behaviour of the liquidity premium when the time to maturity $t$ and the spot price vary

\begin{figure}[!ht]
\begin{center}
   \includegraphics[height=18em]{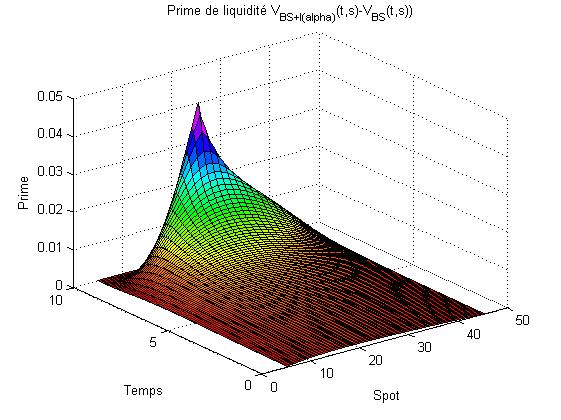}
   \caption{Digital liquidity premium - $T=10$, $K=25$, $\sigma=0.5$, $\epsilon=0.1$, $\ell=1$}
   \end{center}
\end{figure}
\noindent Qualitatively, the liquidity premium behaves as in the Call case. However, as expected the effects of illiquidity are even stronger for ATM options near maturity, since the $\Gamma$ of a digital option explodes faster. Moreover, with our set of parameters, the first order correction to the price is at most $0.04$ for a BS price of $0.21$, which means that the hedge against liquidity risk is much more expensive in the case of a digital option, for a same level of liquidity in the market.

\paragraph{Numerical confirmation of the expansion order}
We represent below the liquidity premium for a fixed value of the spot when the parameter $\eps$ varies with a logarithmic scale.
\begin{figure}[!ht]
\begin{center}
   \includegraphics[height=17em]{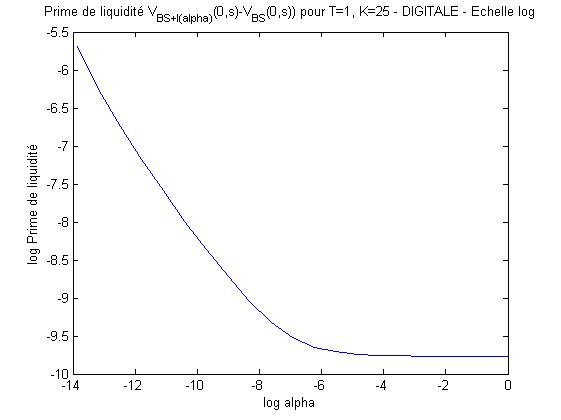}
   \caption{$\log\left(V^\eps-v^{BS}\right)$ - $T=1$, $K=25$, $s=15$, $\sigma=0.5$, $\epsilon=0.1$, $\ell=1$}
   \end{center}
\end{figure}

\vspace{40em}
\noindent For small values of $\eps$ we observe the expected linear behaviour of $\log\left(V^\eps-v^{BS}\right)$. The slope of the above curve is roughly equal to $1/2$ (the exact value here is $0.54$), which is close to our minimal value of $2/5$. The numerical results suggest that the true expansion order lies in the interval $[2/5,1/2]$.

\vspace{0.5em}
\noindent It is also important to realize the financial implications of our results. We just have highlighted the fact that the first order effect exhibits a phase transition for discontinuous payoff, in the sense that derivative securities of the type of digital options induce a cost of illiquidity which vanishes at a significantly slower rate than the continuous payoff case. This means that derivative with discontinuous payoff are more rapidly affected by the illiquidity cost.

\vspace{1em}

\begin{paragraph}{Acknowledgements}
The authors whish to thank Reda Chhaibi for letting them use his Matlab code for the numerical resolution of the PDE \reff{PDE}. 
\end{paragraph}

\newpage
\begin{appendix}
\section{Technical Proofs}
\proof[Proof of Proposition \ref{prop.ineq}]
\noindent We start by proving the inequality for $v^{BS,\alpha}_{ss}$. By dominated convergence, it is clear that $sv^{BS,\alpha}_{ss}$ goes to $0$ when $s$ approaches $0$ or $+\infty$. Hence for $\alpha\neq 0$, it also converges to $0$ when $\tau$ tends to $0$. Thus $sv^{BS,\alpha}_{ss}$ is less than a constant $C_\alpha$ independent of $s$ and $\tau$. However, when $\alpha$ tends to zero, we obtain the classical expression of the $\Gamma$ of a call option 

\begin{equation}
v^{BS}_{ss}(t,s)=\frac{e^{-\frac12d_1(s,K,\tau)^2}}{s\sigma\sqrt{2\pi\tau}},
\label{gamma.call}
\end{equation}
which is known to explode only when $s=K$ and $\tau\rightarrow 0$. Therefore, to understand the dependence in $\alpha$ of $C_\alpha$, we only have to study the behaviour of $sv^{BS,\alpha}_{ss}$ when $s=K$ and when both $\alpha$ and $\tau$ go to $0$. 

\vspace{0.8em}
\noindent Let us therefore take $\alpha=\epsilon^a$ and $\tau=\eps^b$ with $a$ and $b$ strictly positive numbers. For all $\beta \in [1/2,1]$ we have
\begin{align*}
\tau^{1-\beta}\alpha^{2\beta-1}sv^{BS,\eps^a}_{ss}&=\frac{\epsilon^{(b/2-a)(1-2\beta)}}{\sigma\sqrt{2\pi}}\int_{-1}^{1}\phi(u)e^{- \frac{1}{2} \left(\frac{\sigma\eps^{b/2}}{2}-\frac{\eps^{-b/2}}{\sigma}\log\left(1+ \frac{\eps^au}{K}\right)\right)^2}du
\end{align*}

\noindent Therefore, if $a<b/2$ (i.e. if $\tau$ goes to $0$ faster than $\alpha$) the quantity above always goes to $0$ when $\eps\rightarrow 0$ due to the exponential term. If $a\geq b/2$, the exponential term goes to $1$, but since $\beta \in [1/2,1]$ the above expression has always a finite limit. Hence the inequality for $sv^{BS,\alpha}_{ss}$.

\vspace{0.6em}
\noindent A change of variable and direct calculations imply that, for all $\nu \in [0,1]$, we have
\begin{equation}
\tau^{\frac{1-\nu}{2}}\alpha^{1+\nu}sv^{(1),\alpha}_{ss}(t,s)=\frac{\alpha^{1+\nu}\tau^{-\frac{1+\nu}{2}}}{8\ell \pi s}\int_0^{1}\int_{(-1,1)^2}\frac{\phi(x)\phi(y)\widetilde{h}_{\alpha}(\tau,\tau v,s,K,x,y)}{ \sqrt{ v}(2-v)^{3/2}}dxdydv,
\label{e.ineq}
\end{equation}

\noindent where 
\begin{align*}
\frac{\widetilde{h}_{\alpha}(\tau,v,s,K,x,y)}{h_{\alpha}(\tau,v,s,K,x,y)}&=2+\left(2\delta(\tau,\tau v,s,K)- \frac{\log\left(1+ \frac{\alpha x}{K}\right)+\log\left(1+ \frac{\alpha y}{K}\right)}{\sigma\sqrt{\tau(2-v)}}\right)^2\\
&+\left(2\delta(\tau,\tau v,s,K)-\frac{\log\left(1+ \frac{\alpha x}{K}\right)+\log\left(1+ \frac{\alpha y}{K}\right)}{\sigma\sqrt{\tau(2-v)}}\right)\sigma \sqrt{\tau(2-v)}.
\end{align*}

\noindent Using the same arguments as in the proof of the previous inequality, we can show again that the only problem corresponds to the case where $s=K$ and $\alpha$ and $\tau$ go to $0$. Using the same notations, we have

\begin{align*}
h_{\eps^a}(\eps^b,\eps^b v,s,s,x,y)&=\exp\left(- \frac{\sigma^2\eps^b(1-2v)^2}{4(2-v)}+ \frac{(1-2v)\left(\log\left(1+ \frac{\eps^a x}{K}\right)+\log\left(1+ \frac{\eps^a y}{K}\right)\right)}{2(2-v)}\right)\\
&\times\exp\left(- \frac{\eps^{-b}}{\sigma^2(2-v)}\log\left(1+ \frac{\eps^a x}{K}\right)\log\left(1+ \frac{\eps^a y}{K}\right)\right)\\
&\times\exp\left( - \frac{\eps^{-b}\left(\log\left(1+ \frac{\alpha x}{K}\right)-\log\left(1+ \frac{\alpha y}{K}\right)\right)^2}{2\sigma^2v(2-v)}\right)\\[0.8em]
\frac{\widetilde{h}_{\eps^a}(\eps^b,v,s,s,x,y)}{h_{\eps^a}(\eps^b,v,s,s,x,y)}&=2+\left(\frac{\sigma\eps^{\frac b2}(1-2v)}{\sqrt{2-v}}+ \eps^{-b}\frac{\log\left(1+ \frac{\eps^a x}{K}\right)+\log\left(1+ \frac{\eps^a y}{K}\right)}{\sigma\sqrt{2-v}}\right)^2\\
&-\left(\frac{\sigma\eps^{\frac b2}(1-2v)}{\sqrt{2-v}}+ \eps^{-b}\frac{\log\left(1+ \frac{\eps^a x}{K}\right)+\log\left(1+ \frac{\eps^a y}{K}\right)}{\sigma\sqrt{2-v}}\right)\sigma \sqrt{2-v}\eps^{\frac b2}.
\end{align*}

\noindent Therefore, if $a<b/2$, $\widetilde{h}_{\eps^a}$ always goes to $0$. Otherwise, the integral has a finite limite but since $\nu \in [0,1]$ and $a\geq b/2$, the expression in (\ref{e.ineq}) has a finite limit. This proves the second inequality.
\ep

\vspace{2em}
\proof[Proof of Proposition \ref{prop.exp}]
\noindent The first result is straightforward and only uses the fact that the function $\phi$ is symmetric, which allows us to get rid off the odd terms in the expansion. For the second one, we directly calculate that

\begin{align*}
v^{\scriptstyle(1),\alpha}&=\int_0^\tau\int_{-1}^1\int_{-1}^1\phi(x)\phi(y)\frac{e^{-\delta^2- \frac{\alpha^2(x-y)^2}{4K^2\sigma^2v(1- \frac{v}{2\tau})}+o(\alpha^2)}}{8\pi\ell\sqrt{v(2\tau-v)}}dxdydv\\
&+\alpha\int_0^\tau\int_{-1}^1\int_{-1}^1\phi(x)\phi(y)\frac{e^{-\delta^2- \frac{\alpha^2(x-y)^2}{4K^2\sigma^2v(1- \frac{v}{2\tau})}+o(\alpha^2)}\delta}{8\pi\ell K\sigma \sqrt{v}(2\tau-v)}(x+y)dxdydv\\
&+\alpha^2\int_0^\tau\int_{-1}^1\int_{-1}^1\phi(x)\phi(y)\frac{e^{-\delta^2- \frac{\alpha^2(x-y)^2}{4K^2\sigma^2v(1- \frac{v}{2\tau})}+o(\alpha^2)}\left(\scriptstyle2(x+y)^2\delta^2+\sigma \sqrt{2\tau-v}(x^2+y^2)\delta-2xy\right)}{16\pi\ell K^2\sigma^2\sqrt{v}(2\tau-v)^{3/2}} \scriptstyle dxdydv\\
&+o\left(\alpha^2\int_0^\tau\int_{-1}^1\int_{-1}^1\phi(x)\phi(y)\frac{e^{-\delta^2- \frac{\alpha^2(x-y)^2}{4K^2\sigma^2v(1- \frac{v}{2\tau})}+o(\alpha^2)}}{8\pi\ell\sqrt{v(2\tau-v)}}dxdydv\right),
\end{align*}
\noindent where we suppressed the arguments of the functions $v^{(1),\alpha}$ and $\delta$ for notational simplicity.

\noindent Note that all the above integrals are well-defined and finite. Then using dominated convergence and the fact that $\phi$ is symmetric, it is easy to show that

\begin{align*}
v^{(1),\alpha}&=\int_0^\tau\int_{-1}^1\int_{-1}^1\phi(x)\phi(y)\frac{e^{-\delta^2- \frac{\alpha^2(x-y)^2}{4K^2\sigma^2v(1- \frac{v}{2\tau})}+o(\alpha^2)}}{\sqrt{8\pi\ell v(2\tau-v)}}dxdydv\\
&+\alpha\int_0^\tau\int_{-1}^1\int_{-1}^1\phi(x)\phi(y)\frac{e^{-\delta^2}\delta}{8\pi\ell K\sigma \sqrt{v}(2\tau-v)}(x+y)dxdydv\\
&+\alpha^2\int_0^\tau\int_{-1}^1\int_{-1}^1\phi(x)\phi(y)\frac{e^{-\delta^2}\left(\scriptstyle2(x+y)^2\delta^2+\sigma \sqrt{2\tau-v}(x^2+y^2)\delta-2xy\right)}{16\pi\ell K^2\sigma^2\sqrt{v}(2\tau-v)^{3/2}} dxdydv +o\left(\alpha^2\right)\\[0.8em]
&=\int_0^\tau\int_{-1}^1\int_{-1}^1\phi(x)\phi(y)\frac{e^{-\delta^2- \frac{\alpha^2(x-y)^2}{4K^2\sigma^2v(1- \frac{v}{2\tau})}+o(\alpha^2)}}{8\pi\ell\sqrt{v(2\tau-v)}}dxdydv +o(\alpha).
\end{align*}

\noindent Now the first term in the expansion above goes clearly to $v^{(1)}$ as $\alpha$ tends to $0$. Then we have

\begin{align*}
v^{(1),\alpha}-v^{(1)}&=\int_0^\tau\int_{-1}^1\int_{-1}^1\frac{e^{-\delta(\tau,v,s,K)^2}\phi(x)\phi(y)}{8\pi\ell\sqrt{v(2\tau-v)}}\left(e^{- \frac{\alpha^2(x-y)^2}{4K^2\sigma^2v(1- \frac{v}{2\tau})}+o(\alpha^2)}-1\right)dxdydv +o(\alpha).\\
\end{align*}

\noindent Using the change of variable $u=\frac{\alpha|x-y|}{2K\sigma\sqrt{v}}$, the first term above can be rewritten as

\begin{align*}
\frac{\alpha}{8\pi\ell K \sigma} \int_{\frac{\alpha|x-y|}{2K\sigma\sqrt{\tau}}}^{+\infty}\int_{-1}^1\int_{-1}^1\frac{e^{-\delta(\tau,\frac{\alpha^2(x-y)^2}{4K^2\sigma^2 u^2},s,K)^2}\phi(x)\phi(y)|x-y|}{\sqrt{2\tau-\frac{\alpha^2(x-y)^2}{4K^2\sigma^2 u^2}}}\frac{e^{- \frac{u^2}{1- \frac{\alpha^2(x-y)^2}{8\tau K^2\sigma^2 u^2}}+o(\alpha^2)}-1}{u^2}dxdydu.\\
\end{align*}

\noindent A simple application of the dominated convergence and Fubini theorems shows that the above integral (without the $\alpha$ factor) has a finite limit as $\alpha$ approaches $0$ and is given by

\begin{align*}
&\frac{e^{- \frac12d_0(s,K,\tau)^2}}{8\pi\ell K\sigma\sqrt{2\tau}}\int_{-1}^1\int_{-1}^1\phi(x)\phi(y)|x-y|dxdy\int_0^{+\infty}\frac{e^{-u^2}-1}{u^2}du.
\end{align*}

\noindent Since the last integral is equal to $\sqrt{\pi}$, we obtain the second expansion.
\ep

\end{appendix}


\vspace{2em}
\begin{thebibliography}{aa12}

\bibitem{bp} Barles, G. and Perthame, B (1987).
\newblock Discontinuous solutions of deterministic optimal stopping problems.
\newblock {\em Math. Modeling Numerical Analysis}, 21, 557--579.
\bibitem{barles-soner} Barles, G. and Soner, H.M (1998).
Option pricing with transaction costs and a nonlinear
Black-Scholes equation, {\sl Finance and Stochastics}, 2,
369--397.
\bibitem{cjp}
\c{C}etin, U., Jarrow, R. and Protter, P. (2004). Liquidity risk
and arbitrage pricing theory, {\sl Finance and Stochastics}, 8,
311--341.
\bibitem{cjpw}
\c{C}etin, U., Jarrow, R., Protter, P. and Warachka, M. (2006)
Pricing options in an extended Black-Scholes economy with
illiquidity: theory and empirical evidence, {\sl The Review of
Financial Studies}, 19, 493--529.
\bibitem{cr}
\c{C}etin, U. and Rogers, L.C.G. (2006) Modelling liquidity
effects in discrete time, {\sl Math. Finance}, forthcoming.
\bibitem{cst}
Cetin, U., Soner, H.M., and Touzi, N. (2007). Options hedging for small
investors under liquidity costs, preprint.
\bibitem{chsta}
Cheridito, P., Soner, H.M. and Touzi, N. (2005a). The
multi-dimensional super-replication problem under gamma
constraints, {\sl Annales de l'Institute Henri Poincar\'e (C) Non
Linear Analysis}, 22 (5): 633-666.
\bibitem{chstb}
Cheridito, P., Soner, H.M. and Touzi, N. (2005b). Small time path
behavior of double stochastic integrals and applications to
stochastic control, {\sl Annals of Applied Probability}, 15 (4):
2472-2495.
\bibitem{cstv} Cheridito, P., Soner, H.M.,
Touzi, N., and Victoir, N. (2007). Second Order Backward Stochastic
Differential Equations and Fully Non-Linear Parabolic PDEs, {\sl
Communications on Pure and Applied Mathematics},  60 (7): 1081-1110.
\bibitem{cil}
Crandall, M.G., Ishii, H., and Lions, P.L. (1992). User's guide to
viscosity solutions of second order partial differential
equations, {\sl Bull. Amer. Math. Soc.} 27(1), 1--67.
\bibitem{fs89}
Fleming, W.H., and Soner, H.M. (1989).
 Asymptotic expansions for Markov processes with Levy generators.
 {\sl Applied Mathematics and Optimization }, 19(3), 203--223.
\bibitem{fs}
Fleming, W.H., and Soner, H.M. (1993). {\sl Controlled Markov
Processes and Viscosity Solutions}. Applications of Mathematics
25. Springer-Verlag, New York.
\bibitem{fso}
Fleming, W.H., and Souganidis, P.E. (1986).
 Asymptotic series and the method of vanishing viscosity,
 {\sl Indiana University Mathemtics Journal }, 35(2), 425--447.
\bibitem{lsst}
 Lehoczky, J., Sethi, S.P., Soner, H.M., and Taksar, M.I. (1991).
An asymptotic analysis of hierarchical control
of manufacturing systems under uncertainity.
{\sl  Mathematics of Operations Research}, 16(3), 596--608.
\bibitem{sszj}
Sethi, S., Soner,  H.M.,  Zhang, Q., and  Jiang, J. (1992).
Turnpike Sets and Their Analysis in Stochastic Production Planning Problems.
 {\sl  Mathematics of Operations Research}, 17(4), 932--950.
\bibitem{son93}
 Soner, H.M. (1993). Singular perturbations in manufacturing.
 {\sl SIAM J. Control and Opt.} 31(1), 132--146.
 \bibitem{st00}
 Soner, H.M., and Touzi, N. (2000). Super-replication under gamma
 constraints. {\sl SIAM J. Control and Opt.} 39(1), 73--96.
\bibitem{st02}
Soner, H.M., and Touzi, N. (2002). Stochastic target problems, dynamic
programming and viscosity solutions, {\sl SIAM J. Control and
Opt.} 41, 404--424.
\bibitem{st1} Soner, H.M., and Touzi, N. (2002).
Dynamic programming for stochastic target problems and geometric
flows, {\it J.~European Math.~Soc.}, {\bf 4}, 201--236.
\bibitem{st2} Soner, H.M., and Touzi, N. (2007).
The dynamic programming equation for second order stochastic target
problems, preprint.

\end{thebibliography}
\end{document}